\documentstyle[multicol,graphicx,aps,epsf]{revtex}

\begin{document}
\draft
\twocolumn[\hsize\textwidth\columnwidth\hsize\csname@twocolumnfalse%
\endcsname

\title{q-Exponential Distribution in Urban Agglomeration}
\author{ L. C. Malacarne and  R. S. Mendes}

\address{Departamento de F\'\i sica, Universidade Estadual de Maring\'a,
Avenida Colombo 5790,\\ 87020-900,
 Maring\'a-PR, Brazil}
\author{  E. K. Lenzi}
\address{Centro Brasileiro de Pesquisas F\'\i sicas,
R. Dr. Xavier Sigaud 150, \\ 22290-180 Rio de Janeiro-RJ, Brazil }

\date{\today}

\maketitle

\begin{abstract}
Usually, the study of city population distribution  has been
reduced to power laws. In such analysis, a common practice is to
consider cities with more than one hundred thousand inhabitants.
Here, we argue that the distribution of cities for all ranges of
populations can be well described by using a $q$-exponential
distribution. This function, which reproduces the Zipf-Mandelbrot
law,  is related to the generalized nonextensive statistical
mechanics and satisfies an anomalous decay equation.
\end{abstract}

\pacs{PACS number(s): 89.90.+n, 89.65.-s, 05.20.-y}]

In several areas in  nature, besides the complexities, it is
possible to identify macroscopic regularities that can be well
described by simple laws. For example, frequency of words in a
long text\cite{Zipf}, forest fires\cite{5},  distribution of
species lifetimes for North American breeding bird
populations\cite{h2}, scientific citations\cite{6,7}, www
surfing\cite{8}, ecology\cite{8a}, solar flares\cite{8b}, football
goal distribution\cite{mala}, economic index\cite{9}, epidemics in
isolated populations\cite{Rhodes}, among others.

In particular, recently,  the interest in the study of city
population distribution has been increased. Such interest is
related to the analysis of  data and to models that presents the
asymptotic power law behavior\cite{h1,h2,2,3,h23}. However, in
such analysis,  only cities with more than one hundred thousand
inhabitants have been considered. This power law behavior can be
identified in terms of the distribution
\begin{equation}\label{eq1}
N(x)dx \propto x^{-\alpha} dx \;,
\end{equation}
 that gives the number of cities with $x$ and $x+dx$
inhabitants, where $\alpha$ is a positive constant. Another way to
express the same relation is in terms of the relative number (rank
or cumulative distribution) of cities with a population larger
than a certain value x,
\begin{equation}\label{eq1b}
r(x)=\int_x^\infty N(y) dy \propto x^{1-\alpha} \; .
\end{equation}
  By expressing the population $x(n)$ of the cities in descending
order ($x(1)$ being the city with the highest population, $x(2)$
the city with the second highest population, and so on), it
follows from (\ref{eq1b}) that
\begin{equation}\label{eq1c}
x(n) \propto n^{1/(1-\alpha)} \;.
\end{equation}
 The
plot of $x(n)$ on a double logarithmic scale is called a ``Zipf
plot"\cite{Zipf} and leads to a straight line with slop
$1/(1-\alpha)$. Note that the Zipf plot (from Eq. (\ref{eq1c}))
and cumulative  plot (from Eq. (\ref{eq1b})) are equivalent,
except when regarding the weight related to the rare (largest)
elements.

\begin{figure}
 \centering
 \DeclareGraphicsRule{ps}{eps}{*}{}
  \includegraphics*[width=9cm, height=6cm,trim=1cm 0cm 0cm 0cm]{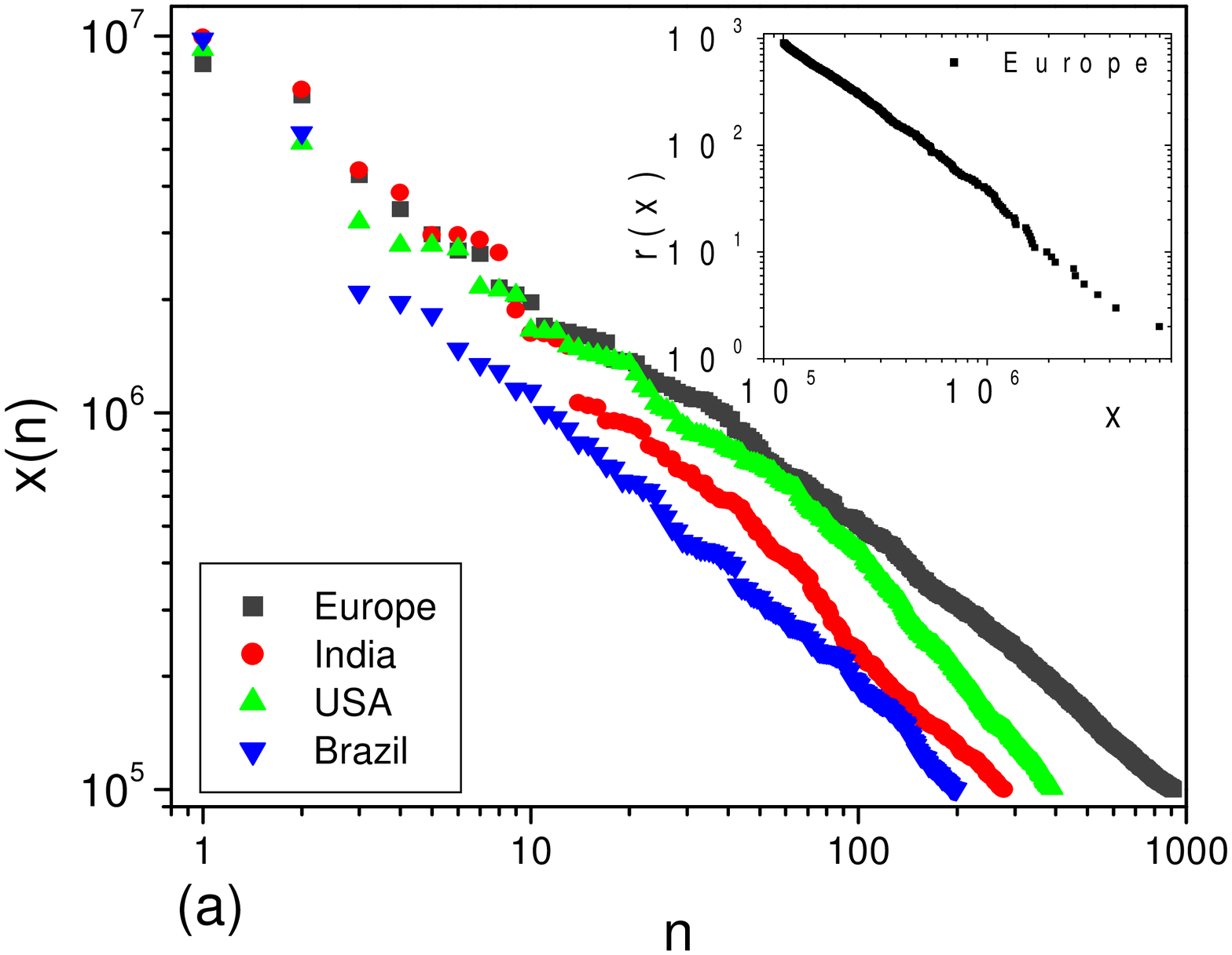}
 \includegraphics*[width=9cm, height=6cm,trim=1cm 0cm 0cm 0cm]{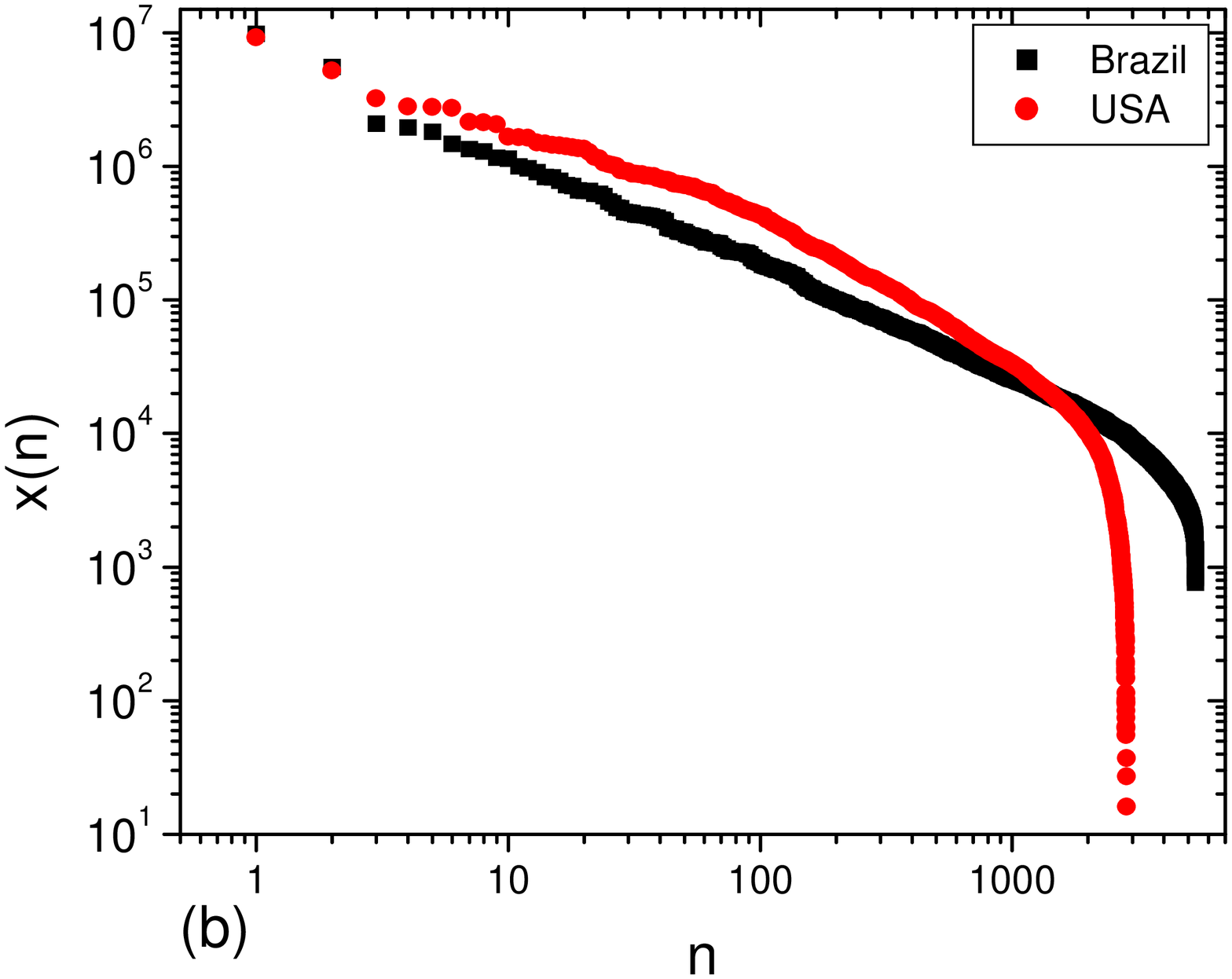}
    \caption{(a) Zipf-plot for cities with population bigger than one hundred
   thousand and, in inset plot, the cumulative Zipf plot to the same cities in Europe.
   (b) Zipf-plot for all cities in USA and Brazil. In the above graphics, $x$ is the
   population of the cities, $n$ is the descending rank and $r$ is the cumulative rank.  }\label{fig1}
\end{figure}

The Zipf plot for cities with more than one hundred thousand
inhabitants\cite{pop} for some countries and Europe is illustrated
in Fig. (\ref{fig1}-a). These graphics enable  us to visualize how
good the power law is at describing  the city population
distribution for large cities.
 In inset plot of Fig. (\ref{fig1}-a) we show the cumulative
 plot for the same cities in Europe. However, there is a little
fraction of  cities with more than a hundred thousand inhabitants.
For instance, these cities represent about $15 \%$ of American
cities and $4\%$ of Brazilian cities. Furthermore, if we take into
account  all cities\cite{usa,brazil} in the country, and by using
the Zipf plot, Fig. (\ref{fig1}-b), we can identify  a notorious
deviation from the asymptotic power law when cities with small
populations are considered.
 Thus, an analysis that considers all
cities is an important task. In  this direction, this work is
dedicated to an empirical analysis of this question.

An alternative approach to incorporate the deviation from
power-law is employed in Ref. \cite{Sornette} by considering the
stretched distribution (Weibull distribution), $N(x)=N_0  x^{c-1}
\exp(-\lambda x^c)$, to fit data of some complex systems. In
particular, for city formation, they also show an adjustment to
cities with population bigger than a hundred thousand inhabitants,
by using  a kind of Zipf plot for $x^c$ versus $\ln (n)$, where
$c$ is an adjustable parameter. However, the Weibull distribution
leads to a poor adjustment for the complete set of data, i.e.,
this distribution give us a satisfactory adjustment only for a
restrict range of data. Furthermore, it is clear that the
stretched function does not lead to an asymptotic straight line in
a log-log plot, i. e., a power law.

On the other hand, Zipf-Mandelbrot law\cite{Mandelbrot},
$N(x)=b/(c+x)^\alpha$ ($b$, $c$, and $\alpha$ all being positive
constants), gives a curvature in a log-log plot, presents an
asymptotic power law behavior and  is normalizable for $\alpha >
1$. In this way, the Zipf-Mandelbrot distribution is a natural
generalization of an inverse power law. This distribution has been
applied in many contexts; in particular, it was recently employed
in the discussion of scientific citations\cite{7} and football
goal distribution\cite{mala}.
 Another important aspect of the Zipf-Mandelbrot's
distribution is that it arises naturally in the context of a
generalized statistical mechanics proposed some years
ago\cite{Tsallis,Curado,Denisov,Tsallis2}. In this framework, the
above distribution is usually rewritten as a q-exponential
function,
\begin{equation}\label{eq2}
N(x)= N_0 \exp_{q'}(-ax) \equiv N_0  [1-(1-q')a x]^{1/(1-q')}\; ,
\end{equation}
where $N_0=bc^{-\alpha}$, $a=\alpha/c$, and $q'=1+1/\alpha$ are
positive parameters. Moreover, the above distribution has been
largely used with $q'<1$ in other contexts\cite{livro}. In this
case, Eq. (\ref{eq2}) is defined equal to zero when $1-(1-q') a
x<0$  in order to overcome imaginary values for $N(x)$. Thus, the
distribution (\ref{eq2}) is equivalent to Zipf-Mandelbrot law only
for $q'>1$ and gives an extension for such law when $q'<1$ is
employed. Note also that $\exp_{q'}(-x)$ reduces to the usual
exponential function, $\exp (-x)$, in the limit $q'\rightarrow 1$.
In addition, Eq. (\ref{eq2}) satisfies an anomalous decay
equation,
\begin{equation}\label{1}
\frac{d}{d x} \left(\frac{N(x)}{N_0}\right) =-a
\left(\frac{N(x)}{N_0}\right)^{q'} \; ,
\end{equation}
independently of the $q'$ value. Since this equation reduces to
the usual decay one in the limit $q'\rightarrow 1$, the parameter
$q'$ can be interpreted as a measure of how anomalous  the decay
is. These aspects put the Zipf-Mandelbrot law in a broad context,
motivating us to employ the generalized Tsallis exponential, Eq.
(\ref{eq2}), instead of the Zipf-Mandelbrot form to study the city
population distribution.

 The cumulative distribution, for $1<q'<1.5$, is
\begin{equation}\label{eq2a}
r(x)= r_0 \left[1-\frac{(1-q)}{q} a x\right]^{1/(1-q)},
\end{equation}
where $r_0 = N_0 q/a$, and $q=(2-q')^{-1}$. Usually, to compare
this cumulative distribution with that obtained from data, it is
employed a log-log plot. Here, we introduce another possible way
to analyze data by using a generalized mono-$\log$ plot based on
the generalized logarithm function, $\ln_q (x)\equiv
(x^{1-q}-1)/(1-q)$. This generalized function arises naturally in
the framework of Tsallis statistics\cite{Tsallis,Curado,Tsallis2}
and reduces to the usual logarithm, $\ln(x)$, for $q\rightarrow
1$. It is easy to verify that the plot of $\ln_q [r(x)]$ versus
$x$ leads to a straight line. So, if the data are well described
by the distribution (\ref{eq2}), we can obtain the $q$-value that
gives the best linear fit in the generalized mono-log plot,
independently of other parameters.

\begin{figure}
 \centering
 \DeclareGraphicsRule{ps}{eps}{*}{}
  \includegraphics*[width=9cm, height=6cm,trim=1cm 1cm 0cm 1cm]{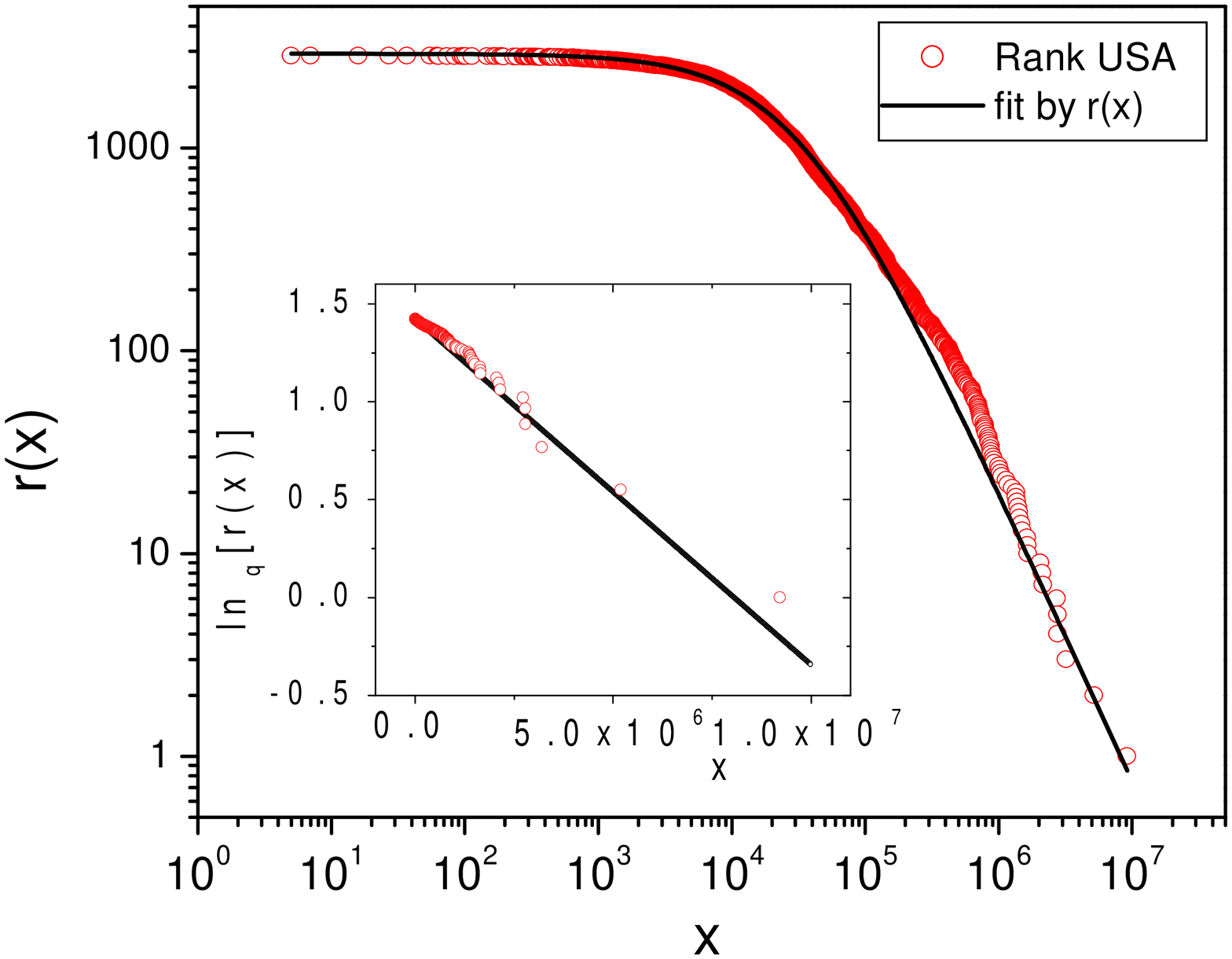}
    \caption{Fit of  cumulative distribution for all cities in USA. The parameters are $q=1.7$,
   $r_0=2919.4$ and
   $a=0.00008$. The coefficient of determination in non-linear fit
   is  $R^2 = 0.99$. Inset plot: generalized mono-log plot for American cities.}\label{usa}
\end{figure}

\begin{figure}
 \centering
 \DeclareGraphicsRule{ps}{eps}{*}{}
 \includegraphics*[width=9cm, height=6cm,trim=1cm 1cm 0cm 1cm]{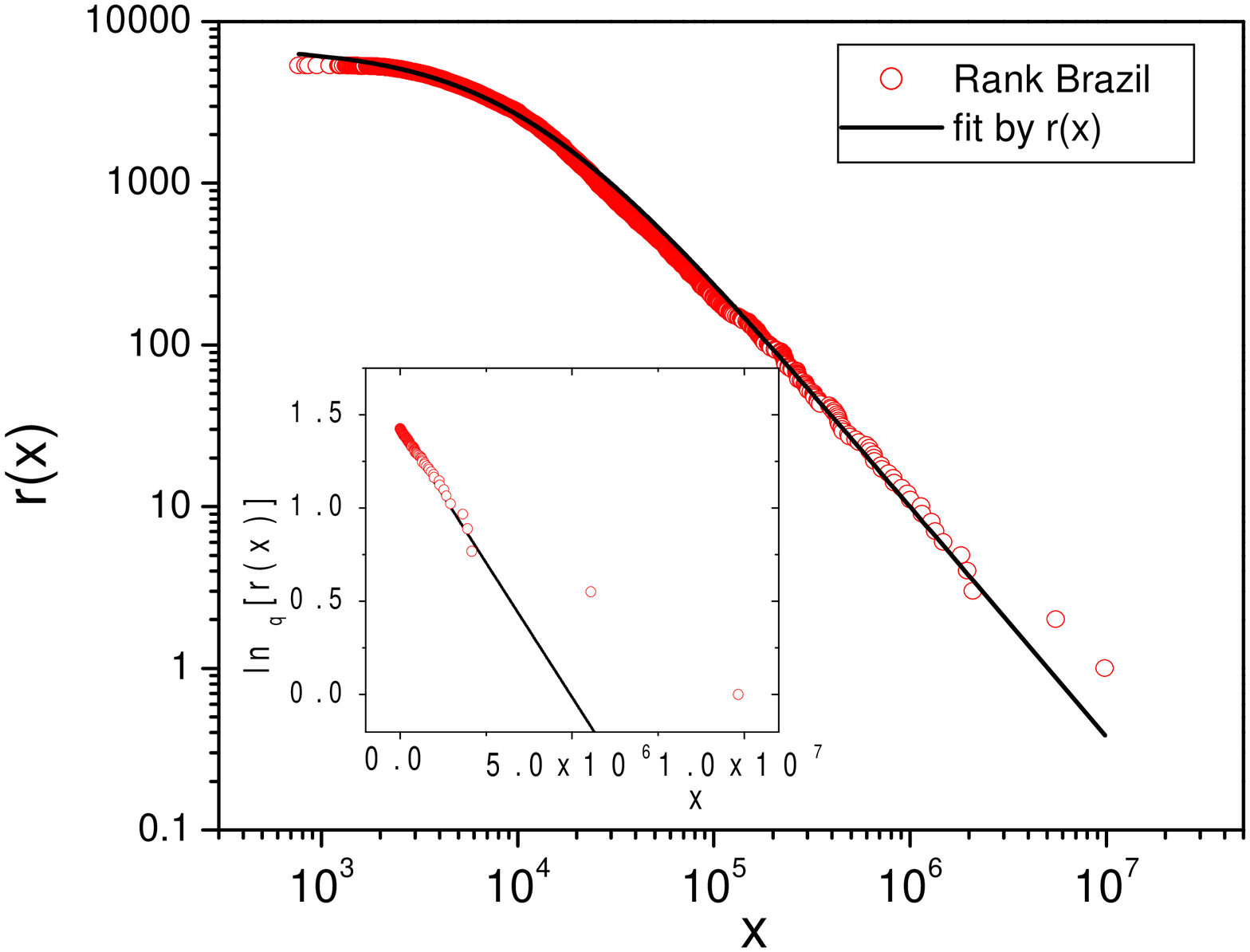}
   \caption{Fit of  cumulative distribution for all cities in  Brazil.
   The parameters are $q=1.7$, $r_0=6968.6$
   and $a=0.00024$.  The coefficient of determination in non-linear fit
   is  $R^2 = 0.99$. Inset plot: generalized mono-log plot for Brazilian cities. }\label{br}
\end{figure}

 Here we used this generalized mono-log plot
analysis and we found that $q\approx 1.7$ gives a good adjustment
to all American  and Brazilian cities. Inset plots of Fig.
(\ref{usa}) and (\ref{br}) show this adjust for American and
Brazilian cities  respectively.  Note that in Fig. (\ref{br}) the
two biggest cities are above the straight line formed by all other
cities. This fact is known as ``king" effect\cite{Sornette,lag},
and occurs because a few cities in some of the countries, by a
specific cause (economic, political, etc), play an irregular
competition to attract people and do not follow the same rule that
most of the cities do. This cities that dominate a region or
country, which is highly centralized, is also referred as
``primate cities" effect\cite{berry}. Of course, this effect can
also be observed if you restrict to cities with more than one
hundred thousand inhabitants. For example, if we consider
countries as England and France, the king effect is related to
London and Paris\cite{Sornette}.

By fixing $q=1.7$, we obtain the other parameters from a
non-linear fit  for the cumulative distribution. This fit is shown
in Fig. (\ref{usa}) for American cities and in Fig. (\ref{br}) for
Brazilian ones.

In order to analyze the agreement between data and the obtained
distribution, beyond what has been visualized in Figs. (\ref{usa})
and (\ref{br}), we calculate the total population $p =
\int_{x_{min}}^\infty x N(x) dx$ and the average  population by
cities by $<x> =\int_{x_{min}}^\infty x N(x) dx /
\int_{x_{min}}^\infty N(x) dx$\cite{limite}. Comparing  $p$ and
$<x>$ with experimental value we obtain  the deviation $\Delta p
\equiv \left[ \frac{p_{data}-p_{model}}{p_{data}}\right] 100\% =
3.9 \%$ for USA cities. Now, considering cities with less than one
hundred thousand inhabitants, we have $\Delta p_{<} =4.6 \%$, that
is better than the one obtained in reference \cite{Sornette} using
the stretched exponential distribution.  For the USA average
population we obtain $\Delta <x> =6.3 \%$. In the Brazilian case,
we obtain $\Delta p= 7.0 \%$ and $\Delta <x> = 9.0 \%$. It is
interesting to remark that the deviations $\Delta <x> $ and
$\Delta p$ could be smaller if the ``king" effect is not present.

In this brief report we show that the population of a country (USA
and Brazil), distributed in its cities, is well described by a
$q$-exponential with $q=1.7$. Thus, this fact indicates a possible
connection among the previous results, Tsallis statistics and
anomalous decay. Furthermore, when one deals with a distribution
that can be adjusted by a $q$-exponential, the generalized
mono-log plot introduced here gives a practical way to determine
the $q$ value, independently of other parameters of the
distribution.

 \acknowledgements We thank  CNPq
(Brazilian Agency) for partial financial suport.



\begin{thebibliography}{10}
\bibitem{limite} In the calculate of total population and average
population, we use the limit of integration,
$\int_{x_{min}}^{x_{max}} dx$, where $x_{min}$ is the population
of the smallest city of the country to be analyzed and $x_{max}$
the largest one. Usually, we can replace $x_{max}\rightarrow
\infty$ since it does not affect substantially the result.
\end{thebibliography}
\end{document}